\pdfoutput=1

\documentclass[11pt]{article}

\usepackage[final]{acl}

\usepackage{times}
\usepackage{latexsym}

\usepackage[T1]{fontenc}

\usepackage[utf8]{inputenc}

\usepackage{microtype}

\usepackage{inconsolata}

\usepackage{graphicx}

\usepackage{xspace}
\usepackage{hyperref}
\usepackage{url}
\usepackage{booktabs}   
\usepackage{xcolor}     
\usepackage{colortbl}   
\usepackage{multicol}   
\usepackage{graphicx}
\usepackage{listings}
\usepackage{xcolor}
\usepackage{amsmath}
\usepackage[most]{tcolorbox}
\newtheorem{Definition}{Definition}

\definecolor{mygreen}{rgb}{0,0.6,0}
\definecolor{mygray}{rgb}{0.5,0.5,0.5}
\definecolor{mymauve}{rgb}{0.58,0,0.82}
\definecolor{mydarkgreen}{rgb}{0.0, 0.5, 0.0}

\newcommand{\tool}{\textsc{VisualCoder}\xspace}

\title{{\tool}: Guiding Large Language Models in Code Execution with Fine-grained Multimodal Chain-of-Thought Reasoning}

\author{
Cuong Chi Le$^{1}$, 
Hoang-Chau Truong-Vinh$^1$, Huy Nhat Phan$^1$, \\
\textbf{Dung D. Le$^2$, Tien N. Nguyen$^3$, Nghi D. Q. Bui$^1$\thanks{Corresponding author.}} \\\\
\normalsize
$^1$FPT Software AI Center, Viet Nam \\
\texttt{\normalsize \{cuonglc4, chautvh, huypn16, nghibdq\}@fpt.com}\\[0.1cm]
\normalsize
$^2$VinUniversity, Vietnam \\
\texttt{\normalsize dung.ld@vinuni.edu.vn}\\[0.1cm]
\normalsize
$^3$University of Texas at Dallas, USA \\
\texttt{\normalsize tien.n.nguyen@utdallas.edu}
}

\begin{document}
\maketitle
\begin{abstract}
Predicting program behavior and reasoning about code execution remain significant challenges in software engineering, particularly for large language models (LLMs) designed for code analysis. While these models excel at understanding static syntax, they often struggle with dynamic reasoning tasks. We introduce \tool, a simple yet effective approach that enhances code reasoning by {\em integrating multimodal Chain-of-Thought (CoT) reasoning} with a visual Control Flow Graph (CFG). By aligning code snippets with their corresponding CFGs, \tool provides deeper insights into execution flows. We address challenges in multimodal CoT integration through a reference mechanism, ensuring consistency between code and its execution path, thereby improving performance in program behavior prediction, error detection, and output generation. Our implementations are available at~\url{https://github.com/FSoft-AI4Code/VisualCoder}.
\end{abstract}

\section{Introduction}
\label{sec:1.intro}

Recent advances in Code-related Large Language Models (LLMs) ~\citep{qwen25coder, codellama, wang2023codet5+, bui2021infercode, nijkamp2023codegen2, lozhkov2024starcoder, stallone2024scaling, to2023better, guo2024deepseek, wei2024magicoder,manh2024codemmlu,huang2024opencoder,muennighoff2023octopack} have pushed the boundaries of complex reasoning tasks, extending to the domains that require an understanding of code and its intricacies. There are diverse approaches aimed at enhancing LLMs' ability. LLMs, while excellent at capturing static patterns and syntax from large code corpora, primarily rely on learned associations rather than direct interaction with the program's execution environment. They struggle with tasks involving dynamic behaviors of programs, such as predicting execution traces, variable values, or runtime errors, because these tasks require precise understanding of runtime context and program states change during execution. They do not inherently simulate code execution, which is necessary for understanding how variables' values vary along program's execution flow. Moreover, LLMs lack the ability to track mutable state or anticipate runtime conditions, leading to difficulties in predicting dynamic behaviors that depends on context-sensitive execution paths.

Recent work has been proposed to enhance the capability of the models in understanding code execution by incorporating Control Flow Graph (CFG) in their reasoning step~\citep{CodeFlow, bieber2020learning, bieber2022static}. It demonstrates that incorporating CFG of given code can significantly improve performance on the code coverage prediction task. However, it utilizes CFGs through graph neural networks rather than directly integrating them into LLM-based reasoning. Despite these advances, the state-of-the-art approaches focus on a single-modality input (i.e., plain code) and has yet to explore the potential of multimodal frameworks for code execution reasoning. While code can be read in a linear fashion, understanding its behavior requires focusing on the non-linear flow of execution, which is visualized more clearly via a~CFG. 

In recent years, Vision Language Large Models (VLLMs)~\citep{openai4, internvl, llava2024}, have made significant progress, showing their potential across a wide range of tasks that involve both visual and textual inputs. These models, which integrate information from multiple modalities, have been successfully applied to tasks like image captioning, visual question answering, and multimodal retrieval. Recent advances in multimodal LLMs, such as Flamingo \cite{10.5555/3600270.3601993}, CLIP \cite{pmlr-v139-radford21a}, and BLIP-2 \cite{10.5555/3618408.3619222}, highlight the benefits of combining visual and textual inputs for enhanced reasoning. Models like LLaVA \cite{NEURIPS2023_6dcf277e} and MiniGPT-4 \cite{zhu2023minigpt4enhancingvisionlanguageunderstanding} show improved performance in multimodal tasks by integrating both visual and textual inputs. Studies have shown that combining visual representations with text significantly improves reasoning, especially in tasks involving complex structures \cite{wei2024gitagraphvisualtextual}. 

In this work, we propose enhancing the program execution reasoning of LLMs by {\em leveraging multimodal reasoning}, combining plain code with visual representations of the corresponding CFG. In our experiments, simply presenting the plain code alongside textual or visual representations of the CFG has poor performance for program execution-related tasks (Section~\ref{sec:5.experiments}). Recent work by ~\citep{Zhang2023MultimodalCR} focuses on improving multimodal reasoning in LLMs using the prominent Chain-of-Thought prompting~\citep{chain_of_thought} with two separate steps: rationale generation and reasoning to produce answers. However, when applied to our multimodal setup of plain code and CFG, that approach suffers from cascading errors, where inaccuracies in rationale generation negatively impact the reasoning and final answers. 

We introduce {\tool}, a simple yet effective {\bf Reference CoT prompting} technique that explicitly links individual lines of code to their corresponding visual elements in the CFG. By making these detailed references, our approach {\em encourages the model to focus on specific connections between the code and its execution flow during multimodal reasoning process}. This technique is expected to improve the LLM’s performance by guiding it to reason more effectively and grounding its reasoning process with more intuitive and informative representation of code graph via imaging, utilizing both the code structure and its execution dynamics.

\section{Related Work}
\label{sec:2.related}
\subsection{Code Large Language Models}
Large Language Models (LLMs) have been widely applied to various code-related tasks, including code understanding, reasoning, and analysis \citep{chen2021evaluating, li2023starcoder, jiang2024mixtral, touvron2023llama, codellama, xu2022systematic, allal2023santacoder, nijkamp2022codegen, phan2024repohyper, to2023better, manh2024codemmlu, phan2024hyperagent}. Early benchmarks \cite{yin2018learning, iyer2018mapping, manh-etal-2023-vault, chen2021evaluating, MBPP-S, hendrycksapps2021} primarily assessed model performance using match-based similarity metrics, which fail to capture deeper reasoning and functional correctness \cite{chen2021evaluating}. Some benchmarks emphasize domain diversity \cite{yin2018learning, iyer2018mapping, manh-etal-2023-vault}, while others, such as HumanEval \cite{chen2021evaluating}, MBPP \cite{MBPP-S}, and APPS \cite{hendrycksapps2021}, focus on specific tasks like function completion or competitive programming. More recent efforts have sought to expand the scope of evaluation, such as ExeDS \cite{huang2022execution}, which targets data science workflows, and ODEX \cite{wang2022execution}, an open-domain evaluation suite. However, these benchmarks primarily assess static code properties and standalone function reasoning, with limited emphasis on execution flow analysis and dynamic behavior prediction—critical aspects for improving LLMs’ ability to reason about code execution.

\subsection{ML-based Fault Localization}
Recent deep learning-based fault localization (FL) techniques like GRACE~\cite{lou2021boosting}, DeepFL~\cite{DeepFL}, CNNFL~\cite{zhang2019cnn}, and DeepRL4FL~\citep{icse21-fl} have significantly advanced FL. GRACE uses a graph-based representation to rank faulty methods effectively. Earlier ML-based FL approaches relied heavily on test coverage~\citep{zheng2016fault}, but struggled to differentiate between failed tests and faulty ones~\citep{TraPT}. Advanced methods like TRANSFER~\citep{meng2022improving} and FixLocator~\citep{fse22} address this by leveraging semantic features and co-fixing detection. CodeT5-DLR~\citep{bui2022detect} uses LLMs for end-to-end bug detection, localization, and repair.



\subsection{Reasoning about Program Execution}

Research into program execution reasoning has advanced through various approaches. They use execution states from constructed programs~\cite{chen2021latent, 
NeXT, shin2018improving} or predict intermediate subgoals to improve search strategies in sequence-to-sequence models~\cite{shi2023exedec}. Another approach trains neural networks to simulate execution, acting as learned interpreters~\cite{bieber2020learning, bieber2022static, CodeFlow}, often relying on specialized architectures to model flows and dependencies. Other works like Scratchpad and Self-Debugging explored LLM-generated reasoning chains, while NExT~\citep{NeXT} uses runtime traces for task-specific rationales.

\section{Motivation}
\label{sec:3.motivation}

\begin{figure*}[!t]
    \centering
        \includegraphics[width=\textwidth]{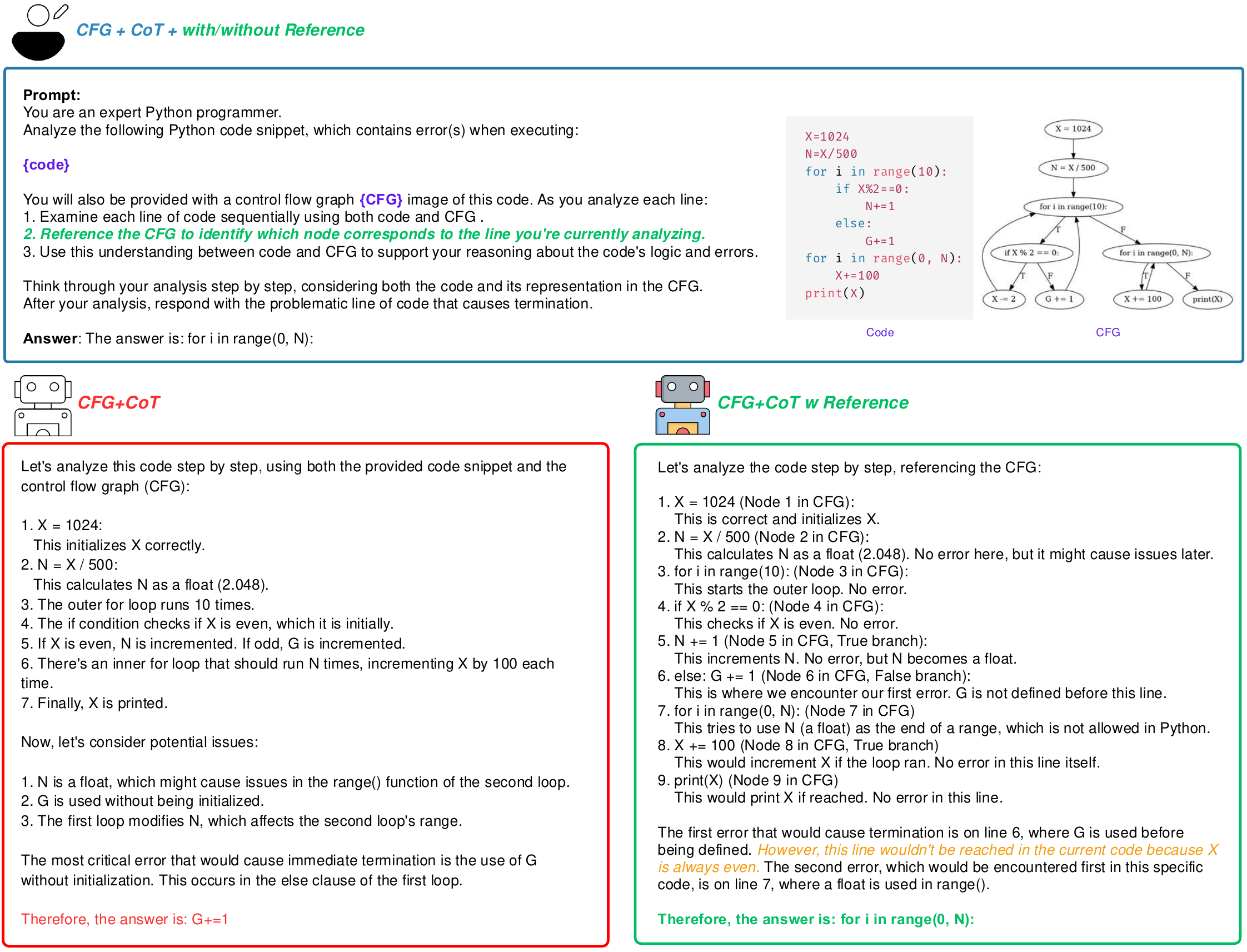}
    \caption{Comparison of Program Execution Reasoning: CFG + CoT w/o Reference vs. CFG + CoT with Reference. With reference, LLM correctly identifies the unreachable node and critical termination point (highlighted in \textcolor{orange}{orange}).}
    \label{fig:cfg_reference_example}
\end{figure*}

Recent advances in Large Language Models (LLMs) have shown promise in tasks like program execution prediction, especially with Chain-of-Thought (CoT) reasoning~\cite{forge24}. However, LLMs still struggle with understanding complex execution flows, such as iterations and conditions. Our results in Table~\ref{tab:code_execution_prediction} (see details later) demonstrate that incorporating \textbf{Control Flow Graphs (CFG)} with source code significantly boosts performance. {\em CFG images provide a visual structure of execution flow}, capturing key control structures like branches and loops. This helps LLMs better understand non-linear execution paths, improving program behavior reasoning.



Choosing the right data representation for CFGs is crucial for helping LLMs understand code execution. To support the use of visual representations, we conducted an experiment comparing the effectiveness of textual vs. visual CFGs. As highlighted in Table~\ref{tab:cfg_image_vs_text} on our experimental results, {\em the models that utilized visual CFG images consistently outperformed those relying on text-based CFG representation}. Our results demonstrate that when models are exposed to CFG images rather than text-based descriptions, their reasoning and prediction accuracy improves substantially. 

Since text-based representations only provide a linear and sequential description of control flow in textual format, they often fall short in capturing the structural complexity of code execution which requires forward-backward reasoning continuously. In contrast, 
{\em the visual modality provides an additional layer of information, allowing the model to better comprehend non-linear code flows, such as loops and branches, which are harder to grasp through sequential textual descriptions alone}. This result is also consistent with the research by Wei {\em et al.}~\cite{wei2024gitagraphvisualtextual}, which emphasizes that incorporating visual representations significantly enhances the reasoning capabilities of multimodal LLMs. Importantly, this result motivates us on the adopting of visual representations that requires deep, non-linear flow of execution reasoning.



Despite the advantages of CFG images, we found that {\em incorporating \textbf{CoT reasoning} into multimodal models is not trivial} and introduces new challenges. Surprisingly, our results in Table~\ref{tab:results} show that adding CoT reasoning alongside CFG images often leads to performance degradation. As seen in Table~\ref{tab:results}, when CoT reasoning was applied to tasks like \textbf{bug detection}, performance dropped for models such as \textbf{Sonnet 3.5} and \textbf{InternVL2-26B}. The models suffer {\em hallucinations}, leading to incorrect reasoning steps. Existing methods, such as the two-stage multimodal Chain-of-Thought (multimodal-CoT) by \cite{Zhang2023MultimodalCR}, attempt to separate rationale generation from answer inference but fail to reason execution on complex code structures.

Let us use an example for illustration. As shown in Figure~\ref{fig:cfg_reference_example}, the \textbf{CFG + CoT} approach fails to capture the critical point in reasoning. As with this approach (see red section), the model incorrectly identifies the termination point within the \textit{else} block (\texttt{G += 1}), missing the fact that this branch is unreachable. Since \texttt{X} is always even, the \textit{else} block will never be executed.

{\em We hypothesize that the key issue is the model's inability to \textit{align} the code with its corresponding CFG image during reasoning}. Without proper alignment with the CFG, the model misinterprets this unreachable path as a valid termination point, focusing on an irrelevant error. Therefore, we guide the model to refer to each line of code to the corresponding element in the CFG as shown in Figure~\ref{fig:cfg_reference_example} (highlighted in {\color{mydarkgreen}{green}}). Let us call it \textbf{CFG + CoT + Reference} approach, which correctly identifies the unreachable node and termination point. 
Our results (Section~\ref{sec:5.experiments}) also show that the two-stage multimodal-CoT approach in~\cite{Zhang2023MultimodalCR}  is also insufficient for complex coding tasks that involve intricate execution flows.

As illustrated in Figure~\ref{fig:cfg_reference_example}, the \textbf{CFG + CoT + Reference} approach ({\color{mydarkgreen}{green}} section) allows the LLM to correctly identify the critical point: the unreachable nature of the \textit{else} branch. By {\em explicitly referencing the CFG during reasoning}, the model avoids errors in unreachable branches and focuses on the actual critical error—the float \texttt{N} being used in the \texttt{range()} function. {\em This reference mechanism helps the model maintain proper alignment between the visual CFG and the code, leading to more accurate reasoning on program execution at runtime}.

Next, we will provide a detailed explanation of our proposed method, demonstrating how the combination of \textbf{Control Flow Graphs}, \textbf{Chain-of-Thought reasoning}, and a \textbf{Reference Mechanism} ({\bf CFG + CoT + Reference}) addresses these challenges and significantly improves code execution reasoning. We formulate our solution in Section~\ref{sec:4.approach}.
\section{{\tool}: Reference Mechanism}
\label{sec:4.approach}

We propose a method that combines \textbf{Control Flow Graphs (CFG)} with \textbf{Chain-of-Thought (CoT)} reasoning, augmented by a \textbf{Reference Mechanism}, to enhance reasoning on program execution. This approach enables step-by-step evaluation of the code while also cross-referencing control flow points, thereby improving error detection and identifying unreachable or erroneous code paths.

Let the given Python code snippet be represented as a sequence of lines of code:
\begin{equation}
\text{Code} = \{C_1, C_2, \dots, C_n\}
\end{equation}
\noindent where $C_i$ represents the $i$-th line or block of code. Along the code, we provide the corresponding {\em Control Flow Graph (CFG)}, which is defined as:
\begin{equation}
\text{CFG} = (N, E)
\end{equation}
where $N = \{N_1, N_2, \dots, N_m\}$ is the set of nodes, each corresponding to a specific code block, and $E \subseteq N \times N$ is the set of directed edges representing control flow between nodes.

The goal is to condition the Vision Large Language Model that semantically maps each line $C_i$ of the code to its node $N_i$ in the CFG, and utilize this to perform stepwise reasoning.

\subsection{Chain-of-Thought Reasoning (CoT)}

Chain-of-Thought reasoning is implemented by analyzing each instruction on $C_i$ while considering its logical dependencies and its corresponding control flow in the CFG. We define the reasoning process as a recursive function:
\vspace{-4pt}
\begin{equation}
R(C_i) = f(C_i, \{C_1, C_2, \dots, C_{i-1}\}, N_j)
\end{equation}

where $f$ is a function that takes the current line of code, its execution context, and its corresponding CFG node $N_j$.

\subsection{Reference Mechanism}

The \textbf{Reference Mechanism} enhances CoT reasoning by mapping each line of code $C_i$ to its corresponding CFG node, expressed as $M: C_i \mapsto N_j$, where $C_i$ is represented by node $N_j$ in the CFG. To establish this mapping, we guide the model to focus on the relevant CFG node while reasoning about each line of code. This is achieved by reinforcing attention on the corresponding node in the CFG image whenever the model processes its associated code line, as demonstrated in Section~\ref{sec:6.2.attention}. By aligning each line of code with its node in the control flow representation, this mechanism improves the model’s understanding of execution paths, transitions, and dependencies across statements and blocks, rather than treating lines in isolation.

\subsection{CFG + CoT (Without Reference)}
In the \textbf{CFG + CoT} approach, the model reasons about the logic purely based on the sequential structure of the plain code. It analyzes each line and attempts to infer potential errors based solely on the textual content, without actively cross-referencing the CFG. This reasoning process can be defined as:
\begin{flalign}
& p_{\text{no-ref}}(Y | C_1, \dots, C_n, \text{CFG}) \nonumber & \\ 
& = \prod_{i=1}^{n} \mathcal{P}(Y_i | C_1, \dots, C_i, \text{CFG}) \nonumber & \\
& = \prod_{i=1}^{n} \mathcal{P}(Y_i | C_1, \dots, C_i, (N_1, \dots, N_m), E) &
\end{flalign}
Here, the probability of generating the correct reasoning $Y$ for the code is determined by the cumulative probabilities of the reasoning steps at each line of code. However, this method is prone to inefficiency, as it includes all CFG nodes $(N_1, N_2, \dots, N_m)$ in each reasoning step, even when many of those nodes are not directly relevant to the current line of code.

\subsection{CFG + CoT + Reference}

In contrast, the \textbf{CFG + CoT + Reference} approach introduces a structured reference to the CFG during each reasoning step. The reasoning at each line $C_i$ is conditioned not only on the previous code lines but also on the corresponding node in the CFG:
\begin{flalign}
& p_{\text{ref}}(Y | C_1, \dots, C_n, \text{CFG}) \nonumber & \\
& = \prod_{i=1}^{n} \mathcal{P}(Y_i | (C_1, M(C_1)), \dots, (C_i, M(C_i)), E) & \label{eq:ref}
\end{flalign}
Where $M(C_i)$ is the mapped node in the CFG corresponding to the current line $C_i$. By analyzing and referencing the corresponding CFG block for every line of code, the model can maintain consistency between the CFG and the source code.




\subsection{\tool}
There are several ways to achieve the behavior outlined in the CFG + CoT + Reference process, such as fine-tuning, one-shot or few-shot prompting, and more. In our current implementation, we propose a straightforward yet effective approach that can be integrated into any Chain-of-Thought framework without the need for fine-tuning. By introducing a simple instruction, as shown in Figure~\ref{fig:cfg_reference_example} (\textcolor{mydarkgreen}{green line} in the prompt), we expect to guide Vision Language Models to follow the formulation described in Equation \ref{eq:ref}. This approach ensures that the model focuses its reasoning on the relevant CFG node for each line of code, thereby improving its alignment with the control flow. The experimental results in Section \ref{sec:5.experiments}, along with the qualitative analysis in Section \ref{sec:6.1.qualitative} and attention heat map in Section~\ref{sec:6.2.attention}, demonstrate the effectiveness of our method in enhancing program execution reasoning.
\section{Empirical Evaluation}
\label{sec:5.experiments}

\subsection{Better Code Execution Understanding with Control Flow Graph Images}

In this experiment, we aimed to show that providing the LLM with CFG images ({\em no references}) improves its code execution reasoning. Using~the CRUXEval benchmark~\cite{CRUXEval}, we tested models on predicting execution outputs. We~compared the accuracies of three state-of-the-art VLM models—Claude 3.5 Sonnet~\cite{anthropic_claude_2024}, Gemini-1.5-Flash~\cite{reid2024gemini}, and InterVL2-8B~\cite{internvl}—in two settings: 1) plain code, and 2) plain code with its CFG image. The task involved both \textbf{output prediction} (predicting the execution's result) and \textbf{input prediction} (predicting the inputs leading to a specific output). 


For direct comparison with prior work, we used the same prompt format from the original CRUXEval paper~\cite{CRUXEval}. The prompt provided the code and, when applicable, a visual CFG, guiding step-by-step reasoning. Performance was measured using the pass@1 metric, indicating if the models' first predictions were correct. 



\begin{table}[ht]
\centering
\resizebox{0.48\textwidth}{!}{
\begin{tabular}{c l c c}
\toprule
\textbf{Task}&\textbf{Settings} & \textbf{Models} & \textbf{pass@1} \\
\midrule
& Plain code  & Claude 3.5 Sonnet & 79.6 \\
& Plain code + CFG image  & Claude 3.5 Sonnet & \textbf{82.3} \\
& Plain code  & Gemini 1.5 Flash & 68.5  \\
Output Pred. & Plain code + CFG image  & Gemini 1.5 Flash & \textbf{70.0 } \\
& Plain code  & InterVL2-8B & 40.8  \\
& Plain code + CFG image  & InterVL2-8B & \textbf{44.0 } \\
\midrule
& Plain code  & Claude 3.5 Sonnet & 75.2  \\
& Plain code + CFG image  & Claude 3.5 Sonnet & \textbf{84.0 } \\
& Plain code  & Gemini 1.5 Flash & 58.4  \\
Input Pred. & Plain code + CFG image  & Gemini 1.5 Flash & \textbf{68.4 } \\
& Plain code  & InterVL2-8B & 43.6  \\
& Plain code + CFG image  & InterVL2-8B & \textbf{44.4 } \\
\bottomrule
\end{tabular}
}
\vspace{-3pt}
\caption{Execution Prediction Performance Comparison}
\label{tab:code_execution_prediction}
\end{table}

The results in Table~\ref{tab:code_execution_prediction} show that incorporating a CFG image improves model accuracy in two settings. This improvement is consistent across models, showing that CFG enhances the LLMs’ ability to reason about execution flow and predict program behaviors more accurately. This result is consistent with the one reported by Le {\em et al.}~\citep{CodeFlow} in which incorporating CFG of given code can improve performance on code coverage prediction.

\subsection{CFG Images vs Text-Based Descriptions}
\label{sec:image-vs-text}



\begin{table}[ht]
\centering
\resizebox{0.3\textwidth}{!}{
\begin{tabular}{l c c}
\toprule
\textbf{Model} & \textbf{CFG (Text)} & \textbf{CFG (Image)} \\
\midrule
Claude 3.5 Sonet & 60.5 & \textbf{74.0} \\
Gemini 1.5 Flash & 65.3 & \textbf{74.1} \\
InternVL2-8B & 23.2 & \textbf{36.4} \\
\bottomrule
\end{tabular}
}
\caption{Comparison of pass@1 results for CFG in text-based description vs. CFG as Image.}
\label{tab:cfg_image_vs_text}
\end{table}

To assess the impact of visual representations in coding tasks, specifically in \emph{Code Execution Prediction}, we conducted an experiment where LLM models were provided with either Mermaid-format (text-based) (see one example of Mermaid-format in \ref{sec:appendix_cfg}) or image-based CFGs, along with the input, and tasked with predicting the code’s output. The prompt remained the same as the previous experiment, but the models received CFGs as images instead of as texts. Results in Table~\ref{tab:cfg_image_vs_text} show that CFG images significantly boost performance in reasoning tasks, underscoring the value of visual aids in enhancing Multimodal LLMs’ reasoning.



\subsection{{\tool} Multi-modal Reasoning}


{\bf Experimental Setting.} This experiment involved two tasks: \textbf{Program Repair} and \textbf{Fault Localization} (further details are provided in Section~\ref{sec:appendix_details}). For \textbf{Program Repair}, we generated a dataset from LiveCodeBench~\cite{livecodebench}, selecting 400 instances to avoid the saturation seen in simpler benchmarks like MBPP-S~\cite{MBPP-S}. We sampled six solutions for each instance using Claude 3.5 Sonnet and Haiku with a 3:1 ratio, ensuring varied difficulty levels. After filtering out fully correct solutions (\textit{those passing all test cases}) and completely incorrect ones (\textit{those failing all test cases}), we retained solutions that have the correct direction—passing a subset of test cases but containing some errors. We finalized 384 solutions for 173 problems. For \textbf{Fault Localization}, we used the FixEval dataset~\cite{fixeval}, which has about 210 programs with various runtime errors. The prompt used was listed in Section~\ref{sec:appendix_prompt}.

Unlike the previous sections, we selected models with stronger code reasoning abilities to handle the increased complexity of Program Repair and Fault Localization task. Specifically, Claude 3.5 Sonnet was retained for its robust performance, GPT-4o replaced Gemini 1.5 Flash due to its superior capabilities and wider adoption, and InterVL2-26B replaced its 8B version, which struggled with coding tasks, often producing generic or incorrect answers.

We evaluated the models in several configurations: plain code (with/without CoT reasoning), plain code with CFGs (with/without CoT), plain code with execution in-line comments (NeXT~\cite{NeXT}), the two-stage \textbf{Multimodal-CoT} method from~\cite{Zhang2023MultimodalCR}, and {\tool}. To assess our method’s adaptability and efficiency, we integrated it with \textbf{Multimodal-CoT} in the first stage of Rationale Generation. The second stage, Answer Inference, remained unchanged. NeXT is excluded from the Fault Localization task, as it relies on code execution, unsuitable for tasks requiring bug detection without execution.



\noindent {\bf Experimental Results.} Table~\ref{tab:results} provides a detailed comparison of \tool with other baseline methods across multiple settings. Chain-of-Thought (CoT) reasoning generally improved model performance, as seen in the Program Repair task where GPT-4o improved from 38.7\% (plain code) to 40.1\% (with CoT), and InternVL2 increased from 0.4\% to 4.0\%. However, combining CoT with CFG images caused a notable performance drop across all models. For instance, Claude 3.5 Sonnet’s accuracy dropped from 63.0\% to 55.5\%, GPT-4o fell from 40.1\% to 37.6\%, and InternVL2-26B dropped from 4.0\% to 2.1\%. Similar declines occurred in the Fault Localization task. This suggests that while CFGs offer structural insights, {\em integrating them with CoT without proper schemes can confuse the models and reduce accuracy}, a finding consistent with~\cite{Zhang2023MultimodalCR}.


In the Program Repair task, which relies more on logical reasoning than execution-heavy tasks, CFGs proved less useful. Although our method didn’t outperform the highest-performing settings (e.g., plain code without CoT for Claude at 64.1\%), it significantly boosted performance over the plain code + CFG w/ CoT setting. It raised Claude 3.5 Sonnet from 55.5\% to 62.9\%, and GPT-4o from 37.6\% to 41.2\%. InternVL2-26B, which struggled with CFG + CoT (2.1\%), improved to 6.3\% with \tool and 10.7\% when combined with Multimodal-CoT. In some cases, it outperformed methods like NeXT and Multimodal-CoT, with Claude 3.5 Sonnet achieving 62.9\% with \tool, compared to 57.3\% with NeXT and 58.7\% with Multimodal-CoT, showing its capability, even in tasks where CFGs are less central.


\begin{table}[t]
\centering
\resizebox{0.48\textwidth}{!}{
\begin{tabular}{l l c c c }
\toprule
\textbf{Tasks} & \textbf{Settings} & \multicolumn{1}{c}{\textbf{Claude}} & \multicolumn{1}{c}{\textbf{GPT-4o}} & \multicolumn{1}{c}{\textbf{InternVL2}} \\
 & & \textbf{3.5 Sonet} & & \textbf{26B} \\
\midrule
& Plain code \textcolor{red}{w/o} CoT & 64.1 & 38.7 & 0.4   \\
& Plain code \textcolor{blue}{w/} CoT & 63.0 & 40.1 & 4.0  \\
 & Plain code + CFG \textcolor{red}{w/o} CoT & 61.2 & 36.5 &  0.9  \\
Program Repair & Plain code + CFG \textcolor{blue}{w/} CoT & 55.5 & 37.6 &  2.1 \\
& NeXT & 57.3 & 40.7 &  0.0  \\
& Multimodal-CoT & 58.7 & 35.1 &  8.2  \\
& \textbf{\tool} & 62.9 & \textbf{41.2} &  6.3 \\
& \textbf{Multimodal-CoT + \tool} & 60.1 & 38.2 &  \textbf{10.7}  \\
\midrule
& Plain code \textcolor{red}{w/o} CoT & 90.4 & 87.1 &  37.0   \\
& Plain code \textcolor{blue}{w/} CoT & 90.0 & 89.5 & 26.1  \\
 & Plain code + CFG \textcolor{red}{w/o} CoT & 86.1 & 79.4 & 22.3  \\
Fault Localization & Plain code + CFG \textcolor{blue}{w/} CoT & 88.0 & 85.6 &  41.0  \\
& Multimodal-CoT & 90.9 & 87.6 &  52.1  \\
& \textbf{\tool} & \textbf{91.4} & \textbf{90.4} &  47.4  \\
& \textbf{Multimodal-CoT + \tool} & \textbf{92.8} & \textbf{91.9} &  \textbf{53.6}  \\
\bottomrule
\end{tabular}
}
\vspace{-2pt}
\caption{Performance Comparison on Program Repair and Fault Localization Tasks}
\label{tab:results}
\end{table}


In the Fault Localization task, improvements were consistent across settings. In the plain code w/o CoT setting, Claude reached 90.4\%, GPT-4o 87.1\%, and InternVL2 37.0\%. Introducing CoT improved GPT-4o to 89.5\%, while Claude remained at 90.0\%. Adding CFGs led to varied results: Claude dropped to 86.1\%, GPT-4o to 79.4\%, and InternVL2 to 22.3\%. These mixed outcomes suggest that while providing structural insights, CFGs complicate reasoning without proper integration.


As shown in Table~\ref{tab:results}, \tool achieved the highest accuracy for both Claude 3.5 Sonnet (91.4\%) and GPT-4o (90.4\%). When combined with Multimodal-CoT, performance further improved, with Claude reaching 92.8\% and GPT-4o 91.9\%. The biggest gain was for InternVL2-26B, which increased from 41.0\% (CoT with CFG) to 53.6\% with \tool and Multimodal-CoT. These results show that {\em integrating CFGs with CoT reasoning and the Reference Mechanism boosts fault localization, especially when paired with Multimodal-CoT. This also highlights the effectiveness of generating rationale that efficiently leverages both plain code and CFG images}.


\noindent{\bf \tool with Complex Java Repositories.} To further evaluate \tool’s effectiveness on more complex, real-world software defects, we extended our experiments to the Defects4J v1.0~\cite{d4j} benchmark. Defects4J comprises 245 real-world Java bugs from five open-source projects: Chart, Closure, Lang, Math, and Time. For this setting, we used GPT-4o and evaluated its fault localization performance using the \textbf{acc@k} metric, which quantifies the number of bugs where the actual buggy location appears among the top k predictions generated by a tool. Table~\ref{tab:defects4j_results} presents the results, comparing Plain Code (Vanilla), Plain Code + CFG, and Plain Code + CFG + Reference Mechanism as \tool.

\begin{table}[h]
\centering
\resizebox{0.4\textwidth}{!}{
\begin{tabular}{lccc}
\toprule
\textbf{Model} & \textbf{acc@1} & \textbf{acc@4} & \textbf{acc@10} \\
\midrule
Plain code            & 47 & 73 & 90 \\
Plain code + CFG                & 54 & 78 & 95 \\
\tool & \textbf{59} & \textbf{80} & \textbf{97} \\
\bottomrule
\end{tabular}
}
\caption{Fault localization accuracy on Defects4J v1.0.}
\label{tab:defects4j_results}
\end{table}

As shown in Table~\ref{tab:defects4j_results}, \tool consistently outperforms the baselines across all metrics, demonstrating its effectiveness in complex fault localization tasks. The integration of CFGs and the Reference Mechanism significantly enhances localization accuracy, particularly at acc@1, where \tool correctly identifies 12 more bugs than the baseline. These results highlight \tool’s robustness in large-scale Java projects, reinforcing its potential for software debugging.

\begin{figure*}[!ht]
    \centering
    \includegraphics[width=\linewidth]{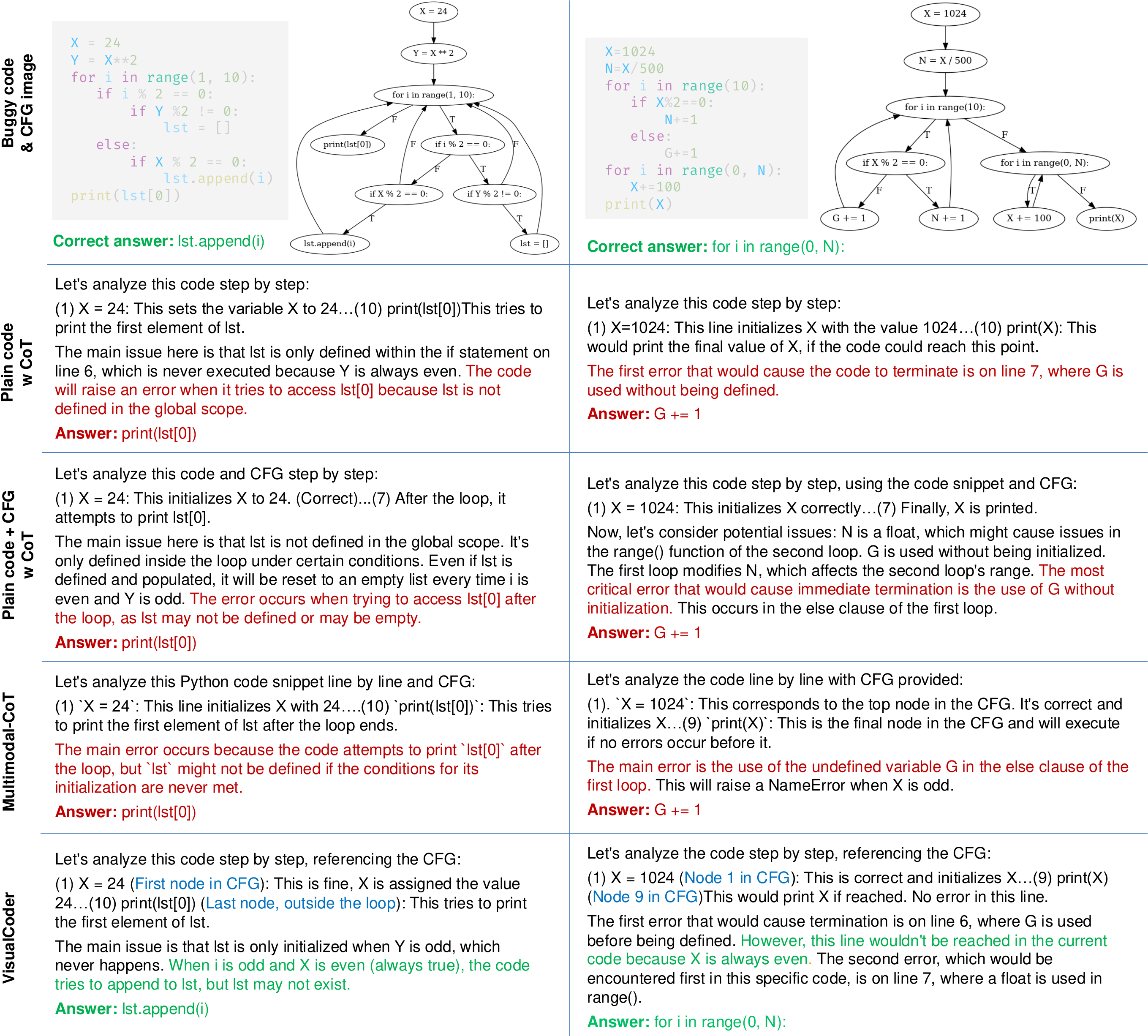}
\caption{Qualitative comparison of reasoning outputs for buggy code using different prompt settings in Claude Sonet 3.5. \textcolor{red}{Red} text indicates where the reasoning fails, \textcolor{mydarkgreen}{green} text highlights correctly identified critical points, and \textcolor{blue}{blue} text in \tool shows the referencing from the plain code to the corresponding nodes in the CFG.}
    \label{fig:qualitative_comparison}
\end{figure*}

\section{Qualitative Analysis}
\label{sec:6.1.qualitative}

Figure \ref{fig:qualitative_comparison} presents two examples of buggy code alongside their corresponding CFGs and the reasoning outputs of Claude Sonet 3.5 under different prompt settings: \textit{plain code with CoT}, \textit{plain code + CFG image with CoT}, and \textit{2-stage prompt of Multimodal-CoT} in~\cite{Zhang2023MultimodalCR}. 


The first three rows of Figure \ref{fig:qualitative_comparison} show Claude Sonnet 3.5’s outputs under different prompt settings, all failing to fully grasp the code's complexity. In the left example (a use-before-initialization error), the model incorrectly identifies \texttt{lst[0]} as the issue, missing the control flow dependencies affecting \texttt{lst}'s initialization. In the right example (unreachable code), it highlights \texttt{G += 1} but overlooks the actual problem: using a float \texttt{N} in the \texttt{range} function. These errors highlight the limitations of plain code reasoning, even with CFG or CoT.

\vspace{-2pt}
The final row shows our approach's result. In the left side, \tool correctly identifies the error by analyzing the CFG and noting the missing connection between \texttt{lst}'s initialization and \texttt{lst.append(i)}. As a result, when the code attempts to append to \texttt{lst}, it triggers a \texttt{NameError} since \texttt{lst} was never initialized, highlighted in \textcolor{mydarkgreen}{green}. Other approaches mistakenly assume \texttt{lst} is reinitialized in each loop iteration, leading to the incorrect conclusion that \texttt{lst[0]} raises an \texttt{IndexError}. Moreover, it uses a reference mechanism (highlighted in \textcolor{blue}{blue}) to link key CFG nodes during reasoning. This helps the model connect execution steps to control flow nodes, a key advantage over methods lacking this explicit referencing.


\vspace{-2pt}
In the example on the right, \tool again shows its advantage by using the CFG to grasp the non-linear control flow. While previous methods failed to identify the incorrect use of the float value \texttt{N} in the \texttt{range} function, it recognizes that the error stems from an unreachable branch of code. The CFG reveals that the \texttt{else} block with \texttt{G += 1} is never executed because \texttt{X} is always even, allowing the model to pinpoint the correct error related to the float value in \texttt{range}. Thus, it accurately identifies \texttt{for i in range(0, N)} as the solution.


{\em These qualitative comparisons highlight our advantage. The red turning points in previous methods indicate breakdowns in reasoning, while the green critical points in our approach's output show how it resolves errors by aligning code with CFG.} 



\begin{figure*}[t]
    \centering
    \includegraphics[width=\linewidth]{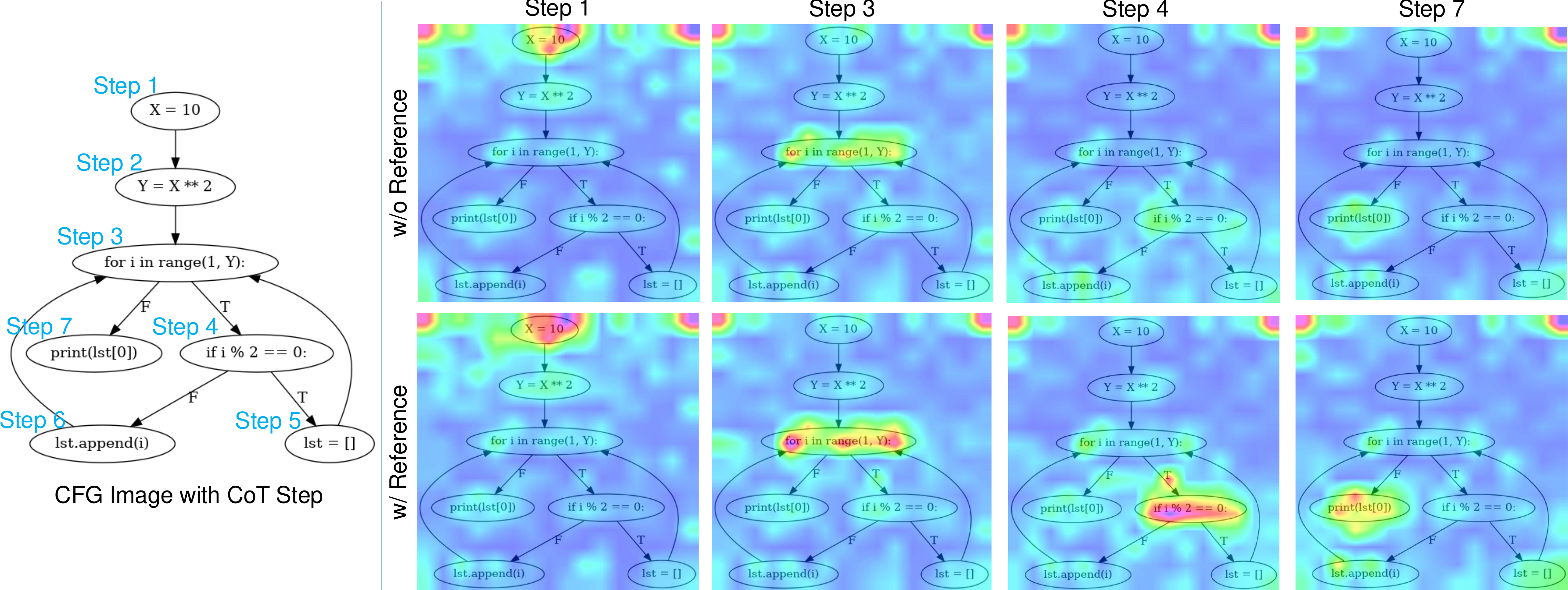}
\caption{Attention Heat Map in CFG Image for each CoT reasoning step.}
     \label{fig:attention}
\end{figure*}

\section{Attention Pattern Analysis}
\label{sec:6.2.attention}

\begin{figure}[ht]
    \centering
    \includegraphics[width=\linewidth]{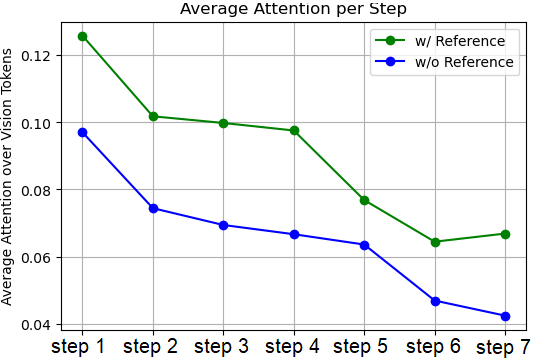}
\caption{Average Attention Score over Vision Token in CFG Image for each CoT reasoning step.}
\vspace{-8pt}
     \label{fig:attention_plot}
\end{figure}

In this section, we aim to analyze to determine whether the Vision-Language Model effectively leverages the CFG images to enhance its reasoning or simply overlooks them during inference. Specifically, we analyze the attention patterns by examining the attention matrices across all heads and layers of the InterVL2-26B model in a specific example (more details in~\ref{sec:appendix_VLM}). Our focus lies on the attention weights associated with the generated rationales and their interactions with visual tokens. These weights provide valuable insights into where and to what extent the model attends to visual tokens during code execution reasoning steps.

As shown in Figure \ref{fig:attention}, the attention maps reveal key differences between the Reference CoT and vanilla CoT approaches. With the Reference CoT, the InterVL2-26B model consistently attends to the relevant nodes in the CFG image at each reasoning step. In contrast, the vanilla approach exhibits a more diffuse attention pattern, occasionally focusing on irrelevant regions, which could contribute to poorer performance in debugging tasks. {\em Figure~\ref{fig:attention_plot} also shows that with Reference Mechanism, VLM more focuses on vision tokens in CFG image for reasoning, leading to better capturing of the alignment between code and CFG image.}
{\em These findings also corroborate our intuition behind Equation \ref{eq:ref}. By incorporating reference mapping, the model adopts a “more focused” attention mechanism for each Chain-of-Thought step, facilitating more precise reasoning on program execution.}

\section{Conclusion}
\label{sec:7.conclusion}


In conclusion, \tool enhances LLMs' reasoning about code execution by incorporating multimodal inputs, specifically control flow graph (CFG) visualizations. Traditional LLMs, while effective at processing static code syntax, struggle to capture dynamic execution behaviors, leading to incorrect predictions and limited reasoning about program flow. By introducing the Reference CoT technique, \tool establishes explicit connections between source code lines and CFG elements, ensuring a structured and interpretable representation of execution logic. This approach reduces reasoning errors, improves alignment between textual and visual execution cues, and enables more accurate program behavior predictions. Our experimental results show that augmenting LLMs with visual CFGs significantly improves performance over text-based CFG descriptions alone, validating our multimodal approach.

\section*{Acknowledgments}
Tien N. Nguyen was supported in part by the US National Science Foundation (NSF) grant CNS-2120386 and the National Security Agency (NSA) grant NCAE-C-002-2021.


\section{Limitations}
\label{sec:8.limitations}

The quality of the CFG plays a crucial role in the performance of our method. If the CFG is incomplete or inaccurate, it can lead to flawed reasoning and missed execution paths. Additionally, we have not tested how well \tool performs when the CFG contains a large number of nodes, which could affect the graph’s resolution and the model’s ability to process fine-grained details. These factors may influence the effectiveness of the reasoning process, particularly in complex programs with extensive control flows.

Another limitation is the lack of clarity on which type of code graph (e.g., CFG, abstract syntax tree, or repository graph) is most suitable for specific coding tasks. While our work focuses on CFGs, other graph representations may be more effective for different types of code reasoning, such as syntax-based analysis or structural relationships in repositories. Identifying the optimal graph type for each task is an area requiring further exploration.

\bibliography{custom}

\appendix

\section{Appendix}
\label{sec:appendix}

\subsection{Control Flow Graph Representation}
\label{sec:appendix_cfg}

\begin{Definition}[Control Flow Graph - CFG]
A Control Flow Graph (CFG) is a graphical representation of the control flow within a program. Nodes in the CFG correspond to basic blocks of code, which may include individual statements or groups of statements that are executed sequentially. The edges between nodes represent the possible transitions or flow of control between these blocks, typically influenced by control structures such as loops, conditional statements (e.g., \texttt{if-else}), or function calls. 
\end{Definition}

In \tool, the CFG serves as a crucial component for visualizing and reasoning about a program’s execution flow. By aligning each code segment with its corresponding node in the CFG, we provide the model with a more structured and intuitive understanding of the dynamic behavior of the program. This enhanced alignment helps in improving code execution reasoning, error detection, and prediction of execution outcomes.

To generate the Control Flow Graphs (CFGs) used in \tool, we adapted code from an open-source repository by \cite{jiang2023cfg}. The modifications made to the original code focused on improving clarity and reducing unnecessary information in the CFG. Specifically, we removed certain function call nodes that did not correspond to any specific line of code, thus eliminating extraneous details that could distract the model. Additionally, we simplified the labels on the edges of conditional branches by replacing the full conditional statements with "T" (True) and "F" (False). 

In addition to the visual representations of Control Flow Graphs (CFGs), we utilize the Mermaid language to provide a text-based representation. The following Mermaid code corresponds to the CFG depicted in Figure~\ref{fig:cfg_reference_example}:

\begin{verbatim}
graph TD
    A["X = 1024"] --> B["N = X / 500"]
    B --> C["for i in range(10):"]
    C --> D["if X % 2 == 0:"] 
    C --> E["for i in range(0, N):"] 
    D --> F["N += 1"] 
    D --> G["G += 1"] 
    E --> H["X += 100"] 
    E --> I["print(X)"] 
    
    D -->|T| F
    D -->|F| G
    E -->|T| H
    E -->|F| I
\end{verbatim}

\subsection{VLM: Patching and Vision Token}
\label{sec:appendix_VLM}
In this section, we describe how we process input contain both text and image using VLM. We use InternVL2-26B to handle the Control Flow Graph (CFG) images. The images are first resized to a resolution of 448x448, then patched and tokenized into 16x16 vision tokens <IMG\_CONTEXT>. These vision tokens are inserted between the language tokens, enclosed by <img> and </img> tags, allowing the model to incorporate the visual information seamlessly into the multimodal input.

After generating the reasoning steps through Chain of Thought (CoT), we map each step to its corresponding language token. To further analyze the model's interaction with the visual context, we calculate the attention scores between each reasoning step and the vision tokens. This process helps us track how much attention the model allocates to the visual tokens at each step. We then plot a heatmap to visualize these attention patterns and compute the average attention score over the vision tokens, providing insight into the model's focus on visual information during the reasoning process.

\subsection{Coding task details}
\label{sec:appendix_details}

This section describes the three coding tasks used to evaluate our approach: \textbf{Input/Output Prediction}, \textbf{Program Repair}, and \textbf{Fault Localization}. These tasks were selected to assess the model’s ability with the help of Control Flow Graphs (CFGs).

In \textbf{Input/Output Prediction}, the model predicts the output of a Python code snippet given specific inputs, or vice versa. This task tests the model's understanding of execution flow, including variable assignments, loops, and conditionals. CFGs are crucial here as they provide a visual representation of the control flow, helping the model trace execution paths more effectively and make accurate predictions.

In the \textbf{Program Repair} task, the model is given a buggy code and must generate a corrected version. CFGs assist the model by highlighting the control flow paths that lead to errors, allowing it to focus on areas where the logic may have broken down. The use of CFGs helps the model better understand the code’s intended execution, leading to more accurate fixes.

The \textbf{Fault Localization} task requires the model to pinpoint the exact lines of code responsible for failures. By leveraging CFGs, the model gains a structured view of the execution flow, enabling it to trace how different parts of the code are interconnected. This visual representation helps the model pinpoint problematic lines more effectively by clarifying control paths and dependencies, offering a deeper understanding of the error's source.

\begin{figure*}[!ht]
\begin{tcolorbox}[colback=gray!5!white,colframe=gray!75!black] 
You are an expert Python programmer.\\ 
Analyze the following Python code snippet, which contains error(s) when executing:\\

\textcolor{orange}{\{code\}} \\

Respond with only the problematic line of code that causes termination. 
\end{tcolorbox} 
\caption{Plain code w/o CoT prompt} 
\label{fig:buggy_w/o_CoT_prompt}
\end{figure*}

\begin{figure*}[!ht] 
\begin{tcolorbox}[colback=gray!5!white,colframe=gray!75!black] 
You are an expert Python programmer.\\
Analyze the following Python code snippet, which contains error(s) when executing:\\

\textcolor{orange}{\{code\}} \\

As you analyze each line: \\

1. Examine each line of code sequentially.

2. Use this understanding to support your reasoning about the code's logic and potential errors.\\

Think through your analysis step by step, and then respond with only the problematic line of code that causes termination. \end{tcolorbox} 
\caption{Plain code w/ CoT prompt}
\label{fig:buggy_w_CoT_prompt} 
\end{figure*}

\begin{figure*}[!ht]
\begin{tcolorbox}[colback=gray!5!white,colframe=gray!75!black]
You are an expert Python programmer.\\
Analyze the following Python code snippet, which contains error(s) when executing:\\

\textcolor{orange}{\{code\}} \\

You will also be provided with a control flow graph (CFG) image of this code. Respond with only the problematic line of code that causes termination.
\end{tcolorbox}
\caption{Plain code + CFG w/o CoT prompt}
\label{fig:buggy_cfg_w/o_CoT_prompt}
\end{figure*}

\begin{figure*}[!ht]
\begin{tcolorbox}[colback=gray!5!white,colframe=gray!75!black]
You are an expert Python programmer.\\
Analyze the following Python code snippet, which contains error(s) when executing:\\

\textcolor{orange}{\{code\}} \\

You will also be provided with a control flow graph (CFG) image of this code. As you analyze each line: \\

1. Examine each line of code sequentially.

2. Use this understanding to support your reasoning about the code's logic and potential errors.\\

Think through your analysis step by step, considering both the code and its representation in the CFG image. After your analysis, respond with only the problematic line of code that causes termination.
\end{tcolorbox}
\caption{Plain code + CFG w/ CoT prompt}
\label{fig:buggy_cfg_w_CoT_prompt}
\end{figure*}

\begin{figure*}[!ht]
\begin{tcolorbox}[colback=gray!5!white,colframe=gray!75!black]
You are an expert Python programmer.\\
Analyze the following Python code snippet, which contains error(s) when executing:\\

\textcolor{orange}{\{code\}} \\

You will also be provided with a control flow graph (CFG) image of this code. As you analyze each line: \\

1. Examine each line of code sequentially.

\textcolor{mydarkgreen}{2. Reference the CFG to identify which node corresponds to the line you're currently analyzing.}

3. Use this alignment to support your reasoning about the code's logic and potential errors. \\

Think through your analysis step by step, considering both the code and its representation in the CFG image. After your analysis, respond with only the problematic line of code that causes termination.
\end{tcolorbox}
\caption{VisualCoder prompt}
\label{fig:visualcoder_prompt}
\end{figure*}

\begin{figure*}[!ht]
\begin{tcolorbox}[colback=gray!5!white,colframe=gray!75!black]
\textbf{Stage 1:}
\begin{tcolorbox}[colback=gray!5!white,colframe=gray!75!black]
You are an expert Python programmer.\\
Analyze the following Python code snippet, which contains error(s) when executing:\\

\textcolor{orange}{\{code\}} \\

You will also be provided with a control flow graph (CFG) image of this code. As you analyze each line: \\

1. Examine each line of code sequentially.

3. Use this understanding to support your reasoning about the code's logic and potential errors. \\

After your analysis, provide a detailed rationale explaining what might be wrong with the code.
\caption{Rationale Generation prompt}
\end{tcolorbox}
\begin{tcolorbox}[colback=gray!5!white,colframe=gray!75!black]
You are an expert Python programmer.\\
Analyze the following Python code snippet, which contains error(s) when executing:\\

\textcolor{orange}{\{code\}} \\

You will also be provided with a control flow graph (CFG) image of this code. As you analyze each line: \\

1. Examine each line of code sequentially.

2. Reference the CFG to identify which node corresponds to the line you're currently analyzing.

3. Use this alignment to support your reasoning about the code's logic and potential errors. \\

After your analysis, provide a detailed rationale explaining what might be wrong with the code.
\caption{Rationale Generation w/ Reference Mechanism prompt}
\end{tcolorbox}
\textbf{Stage 2:}
\begin{tcolorbox}[colback=gray!5!white,colframe=gray!75!black]

You have a Python code snippet containing error(s) and a rationale for the error(s).Code:\\

\textcolor{orange}{\{code\}}\\

Rationale: \textcolor{orange}{\{rationale\}}

Using this rationale, please identify the specific line of code that causes termination. Respond with only the problematic line of code that causes termination."""
\caption{Answer Inference prompt}
\end{tcolorbox}
\end{tcolorbox}
\caption{Multimodal-CoT two-stage prompt}
\label{fig:multimodal-CoT}
\end{figure*}

\subsection{Prompt}
\label{sec:appendix_prompt}

In this section, we present the different prompt configurations used to guide the Vision-Language Model (VLM) in analyzing Python code execution for \textbf{Fault Localization} task. These prompts are designed to test different reasoning approaches, including plain code analysis, Chain-of-Thought (CoT) reasoning, and the integration of Control Flow Graphs (CFGs) to enhance the model's understanding of the code execution flow. The prompt configurations range from basic setups to more sophisticated ones, as shown in Figure~\ref{fig:buggy_w/o_CoT_prompt},~\ref{fig:buggy_w_CoT_prompt},~\ref{fig:buggy_cfg_w/o_CoT_prompt},~\ref{fig:buggy_cfg_w_CoT_prompt}. Furthermore, we implemented a prompt with the Reference Mechanism, as shown in Figure~\ref{fig:visualcoder_prompt}, which explicitly links the reasoning steps with corresponding CFG nodes, thereby grounding the model’s understanding of the control flow. Finally, Figure~\ref{fig:multimodal-CoT} demonstrates a two-stage prompt that incorporates the Reference Mechanism during the Rationale Generation phase, significantly improving the model's capability in error detection and reasoning.

\end{document}